\documentclass[12pt,a4paper]{article}
\usepackage{jheppub,amsmath, amssymb,slashed,url,bm,textgreek,upgreek}

\usepackage [T1]{fontenc}
\usepackage {calligra}
\usepackage [T1]{pbsi}
\usepackage{CJKutf8} 
\usepackage[colorlinks=true,linkcolor=blue]{hyperref}
\usepackage{amsmath,amssymb,amsfonts} 
\usepackage{graphicx}                 
\usepackage{color}                    
\usepackage{hyperref}                 

\newcommand {\beq}{\begin{eqnarray}}
\newcommand {\eeq}{\end{eqnarray}}

\usepackage{color}

\def \be {\begin{equation}}
\def \ee {\end{equation}}
\def \bea {\begin{eqnarray}}
\def \eea {\end{eqnarray}}
\def \nn {\nonumber}
\def \la {\langle}
\def \ra {\rangle}
\def \rr {\raise.35ex\hbox{\small $\prime$}\kern-.17em{\mbox{\large $\imath$}}}
\def \del {\partial}
\def \dels {\partial\kern-.5em / \kern.5em}
\def \As {{A\kern-.5em / \kern.5em}}
\def \Ds {D\kern-.7em / \kern.5em}

\def \m {\mu}
\def \n {\nu}

\def\frac#1#2{{#1\over #2}}

\begin{document}
\thispagestyle{empty}
\begin{flushright}
{\tt YITP-13-24}
\end{flushright}

\title{\bf Weyl Anomaly Induced Stress Tensors in General Manifolds}

\author{\bf Kuo-Wei Huang}
\affiliation{C. N. Yang Institute for Theoretical Physics,
Stony Brook University, Stony Brook, NY 11794, USA}

\emailAdd{Kuo-Wei.Huang@stonybrook.edu}
\abstract{Considering arbitrary conformal field theories in general (non-conformally flat) backgrounds, we adopt a dimensional regularization approach to obtain stress tensors from Weyl anomalies. The results of type A anomaly-induced stress tensors in four and six-dimensions generalize the previous results calculated in a conformally flat background. On the other hand, regulators are needed to have well-defined type B anomaly-induced stress tensors. We also discuss ambiguities related to type D anomalies, Weyl invariants and order of limit issues.}\
\


\maketitle
\fontsize{12pt}{17.5pt}\selectfont

\section{ Introduction} Conformal (Weyl) anomalies (\cite{Deser:1993yx}, \cite{Duff:1993wm}, \cite{Deser:1976}, \cite{BD}, \cite{BPB}) have been important in conformal field theory, renormalization group flow, entanglement entropy and string theory. The conformal (Weyl) transformation is defined by:
\bea
\bar g_{\m\n}(x)=e^{2\sigma(x)} g_{\m\n}(x)=\Omega^2 g_{\m\n}(x)\ .
\eea
A conformally flat background implies that we can take $g_{\m\n}=\eta_{\m\n}$. Conformal anomalies are also called trace anomalies because of the non-vanishing trace of the stress tensor of a (even dimensional) conformal field theory embedded in a curved spacetime background. The anomaly coefficients (or central charges) show up in the trace of the stress tensor,
\begin{equation}
\label{tracegeneral}
\langle T^\mu_\mu \rangle = \frac{1}{(4\pi)^{d/2}}\left( \sum_j c_{dj} I^{(d)}_j - (-)^{\frac{d}{2}} a_d E^{(d)} + \sum_j d_{dj} D^i J^{(d)}_i \right) \ .
\end{equation}
Here $E_d$ is the Euler density in $d$ dimensions (Type A anomaly). Our convention for
the Euler density is that
\bea
\label{Ed}
E_d = \frac{ 1}{2^{d/2}}
\delta_{\mu_1 \cdots \mu_d}^{\nu_1 \cdots \nu_d}
{R^{\mu_1 \mu_2}}_{\nu_1 \nu_2} \cdots
{R^{\mu_{d-1} \mu_d}}_{\nu_{d-1} \nu_d} \ ,
\eea
and $ I^{(d)}_j $ are independent Weyl invariants (Type B anomalies). In 2D, there are no Weyl invariants. In 4D, there is only one Weyl invariant while in 6D, there will be three Weyl invariants. The last term in \eqref{tracegeneral} denotes the type D anomalies which are total derivatives that could be cancelled by the Weyl variation of local covariant counterterms. 

On the other hand, the main problem when studying any quantum field theory is to determine the renormalized energy momentum tenser (stress tensor). It was shown that the stress tensors of arbitrary conformal field theories in a conformally flat background could be obtained purely from the trace anomalies (\cite{Herzog:2013ed}, \cite{A}, \cite{B}) without the knowledge of a Lagrangian and without supersymmery requirements. The purpose of the present paper is to generalize the results in \cite{Herzog:2013ed} to arbitrary general (non-conformally flat) backgrounds using the dimensional regularization method.

Besides additional calculations needed for obtaining the stress tensors in general backgrounds, there is a conceptual obstacle: When one wants to obtain the stress tensor from type B anomalies via dimensional regularization, a subtle issue regarding a well-defined $n\to d$ limit appears.  In fact, this issue was mentioned in \cite{A} where they argued that dimensional regularization could only work when using conformal flatness.  We will detail this issue and also provide a solution to it in Sec. 3.

The organization of the paper is as follows: In Sec. 2, we define our notation by reviewing the strategy of obtaining the stress tensor in a conformally flat background \cite{Herzog:2013ed}. In Sec. 3.1, we discuss the main issue of having a well-defined dimensional regularization method when the spacetime is not conformally flat. Our main formula will also be given in this section. In Sec. 3.2, we obtain the corresponding stress tensors from type A anomalies in 4D and 6D in general backgrounds. These results generalize the previous results calculated in a conformally flat background (\cite{Herzog:2013ed}, \cite{A}, \cite{B}). In Sec. 3.3, we obtain the 4D type B anomaly-induced stress tensor in general backgrounds. We also discuss the appearance of the term $\sim D^2 R$ from the type B anomaly. We will comment on various ambiguities related to Weyl invariants in Sec. 3.4, where the 4D type D anomaly-induced stress tensor is also given. In the final discussion section, we  compare our 4D results with the literature.
\section{Stress tensors in Conformally Flat Backgrounds}
Here we first review the strategy of obtaining the stress tensors in conformally flat backgrounds (\cite{Herzog:2013ed}, \cite{A}, \cite{B}). Let $Z[g_{\m\n}]$ be the partition function. The effective potential is given by
$
\Gamma[\bar g_{\m\n}, g_{\m\n}]= \ln Z[\bar g_{\m\n}]-\ln Z [g_{\m\n}] 
$.
The expectation value of the stress tensor $\la{T^{\m\n}}\ra$ is then defined by the variation of the effective potential with respect to the metric. 
For a conformally flat background, $\bar g_{\m\n}(x)=e^{2\sigma(x)} \eta_{\m\n}$, we normalize the stress tensor in the flat spacetime to be zero. The (renormalized) stress tensor is given by
\bea
\label{d}
\la{T^{\m\n}(x)}\ra={2\over\sqrt{-\bar g}} {\delta \Gamma [\bar g_{\alpha \beta}] \over \delta \bar g_{\m\n}(x)} \ .
\eea 
It could be shown that the following equation determines the general relation between the stress tensor and the trace anomalies \cite{A}:
\bea
\label{final}
{\delta \sqrt{-\bar g} \la{\bar T^{\m\n}(x)}\ra\over \delta \sigma(x')}=2{\delta \sqrt{-\bar g} \la{\bar T^\lambda_\lambda (x')}\ra\over \delta \bar g_{\m\n}(x)} \ .
\eea
Here we have normalized the stress tensor of flat spacetime to zero. In the scheme with no type D anomalies, we further assume \cite{Herzog:2013ed} that we could always re-write the anomalies as $\sigma-$exact forms using the following identities:
\bea
\label{E}
{\delta\over (n-d) \delta \sigma(x)} \int d^n x'\sqrt{-g}{E_d}(x')&=&\sqrt{-g} {E_d}\ ,\\
\label{W}
{\delta\over (n-d) \delta \sigma(x)} \int d^n x'\sqrt{-g}{
{\cal I}^{(d)}_j}(x')
&=&\sqrt{-g} { I}^{(d)}_j \ .
\eea 
We do not alter $E_d$ in moving away from $d$ dimensions but we alter the form of the $I_j^{(d)}$. We let $\lim_{n \to d} {\cal I}_j^{(d)} = I_j^{(d)}$ where ${\cal I}_j^{(d)}$ continues to satisfy the defining relation
$\delta_\sigma {\cal I}_j^{(d)} = - d \, {\cal I}_j^{(d)}$. We ignore $\lim_{n\to d}$ in \eqref{W} for the simplicity of the expression. The $n$-dimensional Weyl tensor is given by
\bea
\label{nw}
{W^{(n)\mu\nu}}_{\lambda\sigma}\equiv {R^{\mu \nu}}_{\lambda\sigma}
- \frac{1}{n-2} \left[ 2(\delta^\mu_{[\lambda} R^\nu_{\sigma]}
+ \delta^\nu_{[\sigma} R^\mu_{\lambda]} ) +{R \, \delta^{\mu\nu}_{\lambda\sigma}\over (n-1)} \right] \ .
\eea

Factoring out the sigma variation in \eqref{final} and setting the integration constant to zero in flat spacetime, one obtains an intermediate formula
\bea
\label{im}
\lefteqn{
\la{\bar T^{\m\n}}\ra=\lim_{n\to d} {1\over (n-d)} {2\over \sqrt{-\bar g} (4\pi)^{d/2}}
} \\
&&
\times
{\delta\over \delta \bar g_{\m\n}(x)}\int d^n x' \sqrt{-\bar g} \left({\sum_j c_{dj} {\cal I}^{(n)}_j - (-)^{\frac{d}{2}} a_d E_d} \right)|_{\bar g} \ .
\nn
\eea
Following the argument in \cite{Herzog:2013ed} and \cite{A} that the type B anomalies do not contribute to the stress tensors in a conformally flat background due to the fact that there are at least quadratic Weyl tensors defined in the type B anomalies, the stress tensor in a conformally flat background then could be obtained by varying only the Euler density and is given by \cite{Herzog:2013ed}: 
\bea
\label{cff}
\langle \bar T^\mu_\nu \rangle = - \frac{a_d}{(-8 \pi)^{d/2}}
\lim_{n \to d} \frac{1}{n-d}
\Big[{R^{\nu_1 \nu_2}}_{\mu_1 \mu_2} \cdots
{R^{\nu_{d-1} \nu_d}}_{\mu_{d-1} \mu_d} \, \delta^{\mu_1 \cdots \mu_d \mu}_{\nu_1 \cdots \nu_d \nu}\Big]|_{\bar g} \ ,
\eea
where the factor of $(n-d)$ would be eliminated when using the conformal flatness condition by contracting with $\delta^{\nu_j}_{\mu_j}$.

\section{Generalization to Non-Conformally Flat Backgrounds}
\subsection{General Strategy} 
Using \eqref{im}, we saw in \eqref{cff} that the $1\over {n-d}$ could be cancelled by a factor of $(n-d)$ in the conformally flat case after the metric variation. Thus, the limit ${n\to d}$ is well-defined. However, for general (non-conformally flat) backgrounds, we need to check that the limit ${n\to d}$ can be still well-defined. 
 
In the type A case, we do not have this issue because the type A anomaly is a topological quantity. \footnote{One might think the fact that the variation of the Euler density with respect to the metric vanishes in integer dimensions would imply type A anomalies must give terms all proportional to $(n-d)$ to some positive powers after the metric variation.  But it is not true. Let's take 4D as an explicit example: In 4D, the metric variation on the type A anomaly in fact would give additional terms that are not proportional to $(n-4)$:
\bea
\sim (g^{ab} W_{cdef} W^{cdef} - 4 W^{acde} W^b_{~cde})+{\cal O}(n-4)\ .
\eea
In 4D only, the above expression vanishes as an identity. Hence the metric variation of the 4D Euler density indeed vanishes. A similar structure would apply for higher dimensional conformal field theories' type A anomalies.} 
This means that in the type A anomaly part, after the metric variation in \eqref{im}, it always gives us the form $0\over 0$ in the limit $n\rightarrow d$, thus we can adopt L'Hôpital's rule to obtain meaningful results. We will use the following identity for the type A anomalies:
\bea
\label{E2}
{\delta\over (n-d) \delta \sigma(x)} {\cal A}^{(d)}\equiv{\delta\over (n-d) \delta \sigma(x)} \Big[\int d^n x'\sqrt{-g}{E_d}(x')\Big]=\sqrt{-g} {E_d}\ .
\eea

In the type B case, we will need a regulator to have a well-defined limit ${n\to d}$. (Notice that type B anomalies are generally not invariant under the metric variation.) Let us consider the following identities:
\bea
\label{I22}
{\delta\over (n-d) \delta \sigma(x)} {\cal B}^{(d)}_i\equiv{\delta\over (n-d) \delta \sigma(x)} \Big[\int d^n x'\sqrt{-g}{
{\cal I}^{(d)}_j}(x')- \int d^{d} x'\sqrt{-g}{
{I}^{(d)}_j}(x')\Big]=\sqrt{-g} {I}^{(d)}_j\ ,\nn\\
\eea where we add a term that is essentially the type B anomaly in a given dimension, which is by definition a Weyl invariant quantity. The method to get rid of the infinite contribution is as follows: After the metric variation, the parts without the additional term in \eqref{I22} could be written symbolically as
\bea
\lim_{n\to d}\Big\{{1\over (n-d)}[(n-d)f^{(n)}(R,W)+g^{(n)}(R,W)]\Big\}\ .
\eea
Then the function $g(R,W)$ that causes the infinite contribution now will be combined with the additional term's contribution: $-{1\over (n-d)}[g^{(d)}(R,W)] $. Treating the additional term as a regulator, we now could safely use L'Hôpital's rule
\bea
\lim_{n\to d}{g^{(n)}(R,W)-g^{(d)}(R,W)\over (n-d)}=\lim_{n\to d}{d\over dn}\Big[{g^{(n)}(R,W)}\Big]\ .
\eea
Thus, the stress tensors from the type B anomalies contain the following two finite parts:
\bea
f^{(d)}(R,W) +\lim_{n\to d}{d\over dn}\Big[{g^{(n)}(R,W)}\Big]\ .
\eea Notice that one only needs to add the regulator for type B anomalies and the additional term will not affect the numerical results (since its derivative with respect to $n$ is zero); The regulator is introduced to have a L'Hôpital's rule method.  

The fact that the regulator is needed for a well-defined effective action of the type B anomaly agrees with \cite{Deser:1993yx}, but here we use a different kind of effective action that is given by re-writing trace anomaly as  a $\sigma$-exact form.

Let us now express the full formula more precisely. Denote
\bea
{\cal K}_g={\delta\over \delta g_{\m\n}(x)}\left({\sum_j c_{dj} {\cal B}^{(d)}_j - (-)^{\frac{d}{2}} a_d {\cal A}^{(d)}} \right)_g\ .
\eea
Then we factor out the sigma variation (from \eqref{final}) to get
\bea
\label{formula}
&&\sqrt{-\bar g}\la{\bar T^{\m\n}}\ra-\sqrt{-g}\la T^{\m\n}\ra=\lim_{n\to d} {1\over (n-d)} {2\over (4\pi)^{d/2}}{\cal K}_{\bar g}
-\lim_{n\to d} {1\over (n-d)} {2\over (4\pi)^{d/2}}{\cal K}_{ g}\ .
\eea
We further re-write the above expression as
\bea
\label{formulafinal}
\delta \la{ T^{\m\n}}\ra\equiv \la{\bar T^{\m\n}}\ra-\Omega^{-d}\la T^{\m\n}\ra=\lim_{n\to d} {1\over \sqrt{-\bar g} (n-d)} {2\over (4\pi)^{d/2}}
{\cal K}_{\bar g}-\Omega^{-d}\Big[.....\Big]|_{\bar g\to g}\ ,
\eea where
\bea
\Big[.....\Big]|_{\bar g\to g}\equiv\lim_{n\to d} {1\over\sqrt{- g} (n-d)} {2\over (4\pi)^{d/2}}{\cal K}_{g}\ ,
\eea
which simply denotes the same curvature tensor forms but only with $\bar g$ replaced by $g$. \eqref{formulafinal} is the main formula that we will be using in the following sections. 

\subsection{Type A}

In the 4D case, we obtain
\bea
\label{4da}
\delta\la T^{ab}\ra^{(A)} &=&\la T^{ab (A)}\ra({c. f})|_{\bar g} -{a_4\over (4\pi)^2} \Big[ 4R^{cd}W^{a~b}_ {~~c~ d} + \lim_{n\to 4}{1\over (n-4)} (g^{ab} W_{cdef} W^{cdef} - 4 W^{acde}W_{bcde})\Big]|_{\bar g} \nn\\
&&-\Omega^{-4}\Big[.....\Big]|_{\bar g\to g}\ ,
\eea
where $(c.f)$ denotes the conformally flat case. The 4D stress tensor in a conformally flat background is given by (\cite{Herzog:2013ed}, \cite{A}, \cite{B}) 
\bea
\la T^{ab}\ra^{(A)} ({c.f}) ={-a_4\over(4\pi)^2} \Big[g^{ab} \Big({R^2\over 2}-R^2_{cd} \Big)+2R^{ac}R^{b}_c-{4\over3}RR^{ab}\Big] \ .
\label{fourDresult}
\eea  Notice that \eqref{4da} is obtained by rewritting Riemann tensors into Weyl tensors in order to factor out the $(n-4)$ factors. After rewritting  Riemann/Weyl tensors into Weyl/Riemann tensors, we should treat the remaining tensors as dimension-independent variables. The topological nature of the type A anomalies implies that we can use the L'Hôpital's rule on $\lim_{n\to 4}{1\over (n-4)} (g_{ab} W_{cdef} W^{cdef} - 4 W_{a cde}W_{b}^{cde})$, which gives zero. 
Thus, the result is
\bea
\label{A}
\delta\la T^{ab}\ra^{(A)}_{n=4} &=&\Big[\la T^{ab (A)}\ra({c. f})-{a_4\over (4\pi)^2} 4R^{cd}W^{a~b}_ {~~c~ d}\Big]|_{\bar g} -\Omega^{-4}\Big[.....\Big]|_{\bar g\to g}\ ,
\eea
where the extra term $\sim R^{cd}W^{a~b}_ {~~c~ d}$ vanishes once traced. This result computed in a new way agrees with \cite{Page1}.  Let us now consider order of limit issues. In this 4D type A case, we have
\bea
\label{tr4}
[\lim_{n\to 4}, Tr] \delta\la T^{ab}\ra^{(A)}  =-{a_4\over (4\pi)^2}( I^{(4)}|_{\bar g} -\Omega^{-4}I^{(4)}|_{g})\ ,
\eea where
\bea
I_1^{(4)} = W^{(n=4)}_{\mu\nu\lambda\rho} W^{(n=4) \, \mu\nu \lambda \rho}\ ,
\eea is the only Weyl invariant in 4D. Note that \eqref{tr4} gives zero because of the nature of $I^{(4)}$ which transforms covariantly. We also have
\bea
[\lim_{n\to4} , \lim_{W\to 0}]\delta\la T^{ab}\ra^{(A)} =0\ ,
\eea since $R^{cd}W^{a~b}_{~~c~ d}$ vanishes in a conformally flat background.

Let us next consider the stress tensor derived from the 6D type A anomaly in general backgrounds. We obtain a new result in 6D that (to our knowledge) was not computed before :
\bea
\label{typea}
&&\delta\la T^{ab}\ra^{(A)}_{n=6}  =\la T^{ab}\ra ^{(A)}({c. f})|_{\bar g} +{a_6\over (4\pi)^3} \Big[{12\over5} R R^{cd}W^{a~b}_{~~c~d}-3R^{de}R^{bc}W^{a}_{~dce}-3R_c^eR^{cd}W^{a~b}_{~d~e}\nn\\
&&+6 R^{bc}W^{adef}W_{cdef}+{3\over2}g^{ab}R^{cd}R^{ef}W_{cdef}-12 R^{cd}W^{aebf}W_{cedf}-{3\over2}R^{ab}W^{cdef}W_{cdef}\nn\\
&&+{27\over 20}g^{ab}RW^{cdef}W_{cdef}-6g^{ab}R^{cd}W_{c}^{~efg}W_{defg}
-{27\over 5}RW^{acde}W^{b}_{~cde}-3R^{ac}R^{de}W^b_{~dce}+\nn\\
&&6R^{cd}W^{a~ef}_{~c}W^{b}_{~def}+6R^{ac}W_{cdef}W^{bdef}+12R^{cd}W^{ae~~f}_{~~~c}W^b_{~edf}\Big]|_{\bar g} -\Omega^{-6}\Big[.....\Big]|_{\bar g\to g}\ ,
\eea
where the 6D stress tensor in a conformally flat background is given by \cite{Herzog:2013ed}
\bea
&&\la T^{\m\n}\ra^{(A)}({c.f}) ={a_6 \over (4\pi)^3} [-\frac{3}{2} R_{\lambda}^\m R_{\sigma}^\n R^{\lambda \sigma} +\frac{3}{4} R^{\m\n} R^{\lambda}_{\sigma} R_{\lambda}^{\sigma}+\frac{1}{2} g^{\m\n} R_\lambda^\sigma R^{\lambda}_{\rho} R^{\rho}_{\sigma} \nn\\
&&+\frac{21}{20} R^{\m \lambda} R^\n_{\lambda} R - \frac{21}{40} g^{\m\n} R_{\lambda}^{\sigma}R^{\lambda}_{\sigma} R- \frac{39}{100} R^{\m\n}R^2+ \frac{1}{10} g^{\m\n} R^3] \ .
\eea
In obtaining $\eqref{typea}$ we have dropped $\lim_{n\to 6} (....)$ part \footnote{We have: $\lim_{n\to 6} {1\over (n-6)}(24 W^{acbd}W_{c}^{~efg}W_{defg}-8g^{ab}W_{c~e}^{~g~h}W^{cdef}W_{dhfg}+2g^{ab}W_{cd}^{~~gh}W^{cdef}W_{efgh}
-12W^{acde}W_{defg}W^{b~fg}_{~c}+48 W^{acde}W_{cgef}W^{bf~~g}_{~~~d}).$} since we have the form $0\over 0$ due to the topological nature of the Type A anomaly, as we did in the 4D case. Regarding the order of limit issue, in this case we find:
\bea
\label{tr6}
[\lim_{n\to 6}, Tr] \delta\la T^{ab}\ra^{(A)} &=&-{a_6\over (4\pi)^3} \Big[\Big(8I^{(6)}_1+2 I^{(6)}_2\Big)|_{\bar g} -\Omega^{-6}\Big(8I^{(6)}_1+2 I^{(6)}_2\Big)|_{g}\Big]\ ,
\eea where $I^{(6)}_1$ and $I^{(6)}_2$ are the first two kinds of 6D Weyl invariant tensors (6D Type B anomaly) given by (\cite{Deser:1993yx},\cite{Erdmenger:1997gy},\cite{Bastianelli:2000hi})
\bea
\label{I1}
I^{(6)}_1
&=&W^{(6)}_{\mu \nu \lambda \sigma}~W^{(6)\nu \rho\eta\lambda}~W^{(6)\mu\sigma}_{\rho~~~~~\eta} \ , \\
\label{I2}
I^{(6)}_2
&=&W^{(6)\lambda\sigma}_{\mu \nu}~W^{(6)\rho\eta}_{\lambda\sigma}~W^{(6)\mu\nu}_{\rho\eta} \ , \\
\label{I6}
I^{(6)}_3
&=&W^{(6)}_{\mu \nu \lambda \sigma}\Big(\Box\delta^{\mu}_{\rho}+4R^{\mu}_{\rho}-{6\over5} R \delta^{\mu}_{\rho}\Big) W^{(6)\rho\nu \lambda \sigma}+ D_\mu J^{\mu} \ .
\eea We see again that \eqref{tr6} is zero because of the nature of $I^{(6)}_1$ and $I^{(6)}_2$ that transform covariantly. Finally, similar to 4D, we have
\bea
[\lim_{n\to6} , \lim_{W\to 0}] \delta\la T^{ab}\ra^{(A)} =0\ .
\eea  

\subsection{Type B}

The type B anomaly is not metric variation invariant. We need to introduce the regulator to have the form $0\over 0$ when taking the $\lim_{n\to d}$, as we have mentioned before. Then, after the metric variation, the result from the 4D type B anomaly is given by
\bea
\label{tyb}
&&\delta\la T^{ab}\ra^{(B)}_{n=4} = {c_4\over (4\pi)^2} \Big[-4R^{cd}W^{a~b}_{~c~~d}-g^{ab}R_{cd}R^{cd}+4R^{ac}R^b_{~c}\nn\\
&&-{14\over 9} R R^{ab}+{7\over 18} g^{ab}R+{8\over9}D^{a}D^{b}R-2D^2 R^{ab}+{1\over 9}g^{ab}D^2 R\Big]|_{\bar g} -\Omega^{-4}\Big[.....\Big]|_{\bar g\to g}\ ,
\eea
where we have used L'Hôpital's rule to drop $\lim_{n\to 4} (....)$ part 
\footnote{ We have: $\lim_{n\to 4}{1\over (n-4)}\Big(-2g^{ab}R^{cd}R_{cd}+8R^{ac}R^b_{~c}
-{4\over3}RR^{ab}+{1\over 3}g^{ab}R^2+g^{ab} R^2_{cdef}-4R^{acde}R^b_{~cde}+{4\over3}D^aD^b R-4 D^2 R^{ab}+{3\over 2}g^{ab}D^2 R
\Big)$.}. In this case, we have
\bea
\label{trbb}
[\lim_{n\to 4}, Tr] \delta\la T^{ab}\ra^{(B)}&=& {c_4\over (4\pi)^2}\Big({2\over 3} D^2R|_{\bar g} -\Omega^{-4}~{2\over 3} D^2R|_{g}\Big)\ .
\eea

When the ${2\over3}D^2 R$ term appears in the 4D trace anomaly, one can relate it to an $R^2$ term in the effective action. However, here it shows up as an artifact of dimensional regularization. By taking the $n\to 4$ limit, we have used
\bea
\lim_{n\to 4}\Big[{\delta\over (n-4) \delta \sigma(x)} \int d^n x'\sqrt{-g}
{W^2}(n)(x')\Big]
=\sqrt{-g} { W^2}(4)\ ,
\eea where $W(n)$ is defined in \eqref{nw}. We factored out the $\sigma$ variation, Then the stress tensor was obtained after the metric variation. We found ${2\over3}D^2 R$ in \eqref{trbb} after taking the trace. This process could be formally re-expressed as
\bea
Tr~ {\delta\over \delta g_{\m\n}}\lim_{n\to 4}\Big[{1\over (n-4) } \int d^n x'\sqrt{-g}
{W^2}(n)(x')\Big]\ ,
\eea
which gives
\bea
Tr~ {\delta\over \delta g_{\m\n}}\Big[\Big({1\over (n-4) } \int d^4 x'\sqrt{-g}
{W^2}(4)\Big)|_{n\to4 }+\int d^4 x'\sqrt{-g}
{\del{W^2}(n)\over \del n}|_{n\to4}\Big]\ .
\eea
The divergent first term will be cancelled by the regulator. It is the second term that gives ${2\over3} D^2 R $. \footnote{One can further check that the orders of taking the metric variation and $n\to 4$ expansion commute.}
Therefore, we see that the $ {2\over 3} D^2 R$ has another origin besides adding an $R^2$ term in the effective action. However, it should be stressed that these two ways will give different contributions to the stress tensor via the metric variation, although they both lead to ${2\over 3} D^2 R$ when traced. We notice that there were also several related discussions in AdS/CFT regarding this $ {2\over 3} D^2 R$ term. For instance, \cite{Nakayama:2012jv} discussed this term on page 5 in the context of the holographic c-theorem. \cite{Balasubramanian:1999re} mentioned this kind of ambiguity on page 16. In \cite{Casini:2011kv}, they included the $ {2\over 3}  D^2 R$ term on page 30 to study entanglement entropy.

In 6D, there are three kinds of type B anomalies so that three regulators are needed. One can derive the corresponding transformed stress tensors following the same method we developed here. But the results will be very lengthy so that we do not present then here. Moreover, we will soon comment on ambiguities related to the type B anomalies in the following sections.

\subsection{Type D and Ambiguities}

The type D anomalies give the first kind of arbitrariness in the formulation. In 4D, there is only one kind of the type D anomaly given by:
\bea
\la T^{\mu}_{\mu}\ra ^{(D)}={\gamma\over (4\pi)^2} D^2 R\ ,
\eea
where $\gamma\equiv d_4$ represents the corresponding type D central charge. This anomaly can be generated by using the following identity
\bea
{\delta\over (n-4)(4\pi)^2 \delta \sigma(x)} \Big[\int d^n x'\sqrt{-g} (n-4){-\gamma\over 12}R^2(x')\Big]&=&{\gamma\over (4\pi)^2} D^2 R \ .
\eea
Obviously, there is no $n\to d$ problem here. The stress tensor corresponding to this anomaly is therefore given by the metric variation on the $R^2$ term. We have
\bea
\label{typed}
&&\delta \la T^{ab} \ra ^{(D)}_{n=4} =-{\gamma\over 6 (4\pi)^2} \Big(2D^aD^b R-2g^{ab}D^2 R-2RR^{ab}+{1\over 2}g^{ab}R^2\Big)|_{\bar g}-\Omega^{-4}\Big[.....\Big]|_{\bar g\to g}\ . \nn\\ 
\eea Since one could introduce a counterterm in the effective action to cancel this anomaly, this contribution is arbitrary. In this paper, we will not consider results of stress tensors derived from the 6D type D anomalies, which would presumably lead to lengthy expressions. We refer readers to \cite{Bastianelli:2000rs} for the expressions of all possible type D anomalies in 6D.

Going back to the case of 4D type B anomaly, \eqref{tyb}, one might ask about $\lim_{n\to 4}$ and $\lim_{W\to 0}$ order of limits issue since we consider $\lim_{W\to 0} \la T^{(B)}_{ab}\ra=0$ under the scheme that the type B central charge does not contribute to the stress tensor in a conformally flat background \cite{Herzog:2013ed}. \footnote{Note that \eqref{tyb} is the result after taking $\lim_{n\to 4}$.  If we instead take $\lim_{W\to 0}$ first, we have symbolically $\lim_{W\to 0} {\delta \over \delta g_{\m\n}} \int W^2$, which simply is already zero because of the squared Weyl tensor. }

Our answer to the above question is that there is no definite contribution to the stress tensor from the type B central charge because of various ambiguities related to Weyl tensors. Recall that the main strategy in the dimensional regularization approach is to re-write the trace anomaly into a $\sigma$-exact term. However, one has some arbitrariness that can be added in the effective action: (1) $(n-4) \times\int d^4 x {\sqrt {-g}}R^2$ with an arbitrary coefficient.  This term only modifies the coefficient of the type D anomaly, which is arbitrary as mentioned before; (2) $\sigma$-variation invariant terms such as $(n-4) \times \int d^4 x {\sqrt {-g}} ~type~ A/B~ anomaly$ with an arbitrary coefficient. But notice that the type A anomaly is topological, so it will not contribute to the stress tensor.

By using the first kind of arbitrariness, it is found that if we instead use the following identity
\bea
\label{new}
&&\lim_{n\to 4}{\delta\over (n-4) \delta \sigma(x)} \Big[\int d^n x'\sqrt{-g}{
{\cal I}^{(4)}_j}(x')- \int d^{4} x'\sqrt{-g}{
{I}^{(4)}}(x')\nn\\
&&- (n-4)\Big( {1\over 18 }\int d^{4} x'\sqrt{-g}
R^2(x')\Big)\Big]=\sqrt{-g} [I^{(4)}_j+{2\over3} D^2R]\ ,
\eea
we could modify \eqref{tyb} by adding contributions from the metric variation on the $R^2$ term. We then have the following 4D result:
\bea
\label{typeb}
&& \delta\la T^{ab}\ra^{(B)}_{n=4} =-4{c_4\over (4\pi)^2}\Big(D_cD_dW^{cadb}+{1\over 2} R_{cd}W^{cadb}\Big)|_{\bar g}-\Omega^{-4}\Big[.....\Big]|_{\bar g\to g}=0\ .
\eea
Note that ${\sqrt {-g}} \Big(D_cD_dW^{cadb}+{1\over 2} R_{cd}W^{cadb}\Big)$ is conformal invariant and traceless. Certainly, in this case, we trivially get
\bea
[\lim_{n\to4} , \lim_{W\to 0}] \delta \la T_{ab} \ra^{(B)}=0 \ ,
\eea  and in this case, we will have the same \eqref{trbb} result. 

Regarding the second kind of arbitrariness, we note that because of the following identity:
\bea
&&- 4\sqrt {-g} \Big(D_cD_dW^{cadb}+{1\over 2} R_{cd}W^{cadb}\Big)= {\delta \over \delta g_{ab}} \int d^4x {\sqrt {-g}} W_{abcd}W^{abcd}\ .
\eea
One could generate the form $\Big(D_cD_dW^{cadb}+{1\over 2} R_{cd}W^{cadb}\Big)$ with an arbitrary coefficient. But since this term transforms covariantly, it always give zero contribution to the transformed stress tensor.

At this moment we would like to make a remark on the orders of taking different limits in the formulation: In \cite{Herzog:2013ed}, we followed the same argument in \cite{A} that the type B anomalies do not contribute to the stress tensors in a conformally flat background because of the (at least) squared Weyl tensors. This implies that \cite{Herzog:2013ed} \cite{A} were actually limited to the order: 
\bea
\lim_{n\to 4}~ \lim_{W\to 0}
\eea
 for the conformally flat case. For the order $ \lim_{W\to 4} ~\lim_{n\to 4}$, one should argue firstly why the $n\to 4$ limit is well-defined then use the argument of the squared Weyl tensors for the conformally flat case. The latter consideration is included in this paper.  In fact, using the order $\lim_{n\to 4} ~\lim_{W\to 0} $ was the hidden reason why ${2\over 3}D^2 R$ in $c(W^2+ {2\over3} D^2 R)$ in the trace anomaly gives a separated contribution to the stress tensor in \cite{A}. In \cite{Herzog:2013ed}, we ignored $c{2\over3}D^2 R$ as the scheme to match with AdS/CFT results. Under the order $\lim_{W\to 4} ~\lim_{n\to 4}$ the regulator is needed since the type B anomaly is not a topological quantity. However, this time we will need $c {2\over 3} D^2 R$ to have a result that vanishes in $W=0$. It might be most natural to adopt the scheme that one always introduces the regulator instead of considering the order $\lim_{n\to d} \lim_{W\to 0}$ on the type B anomaly.

Now we discuss an additional ambiguity by observing the following identity \footnote{Note the basic result $
{\delta \bar g_{\m\n}}=2\bar g_{\m\n}\delta \sigma$
implies ${\delta g_{\m\n}\over \delta \sigma}=0$ by considering a fixed $g_{\m\n}$ with respect to the conformal factor.}:
\bea
{\delta\over \delta \sigma(x)}\Big[{1\over 8 } \int d^4x' \sqrt{-\bar g} \bar W^2(x') \ln \bar g(x')\Big]= \sqrt{-\bar g} \bar W^{2}(x)\ .
\eea
After the metric variation, one obtains 
\bea
\label{k}
&&\delta\la T^{ab}\ra^{(B)}_{n=4}  =-{c_4\over (4\pi)^2} \Big[\Big(D_cD_dW^{cadb}+{1\over 2} R_{cd}W^{cadb}\Big)\ln { g}-{1 \over 4} W^2 g_{ab}\Big]|_{\bar g}-\Omega^{-4}\Big[.....\Big]|_{\bar g\to g}\ ,\nn\\
\eea in contrast to \eqref{typeb}. This case gives
\bea
[\lim_{n\to 4},Tr]\delta\la T^{(B)}_{ab}\ra=0=[\lim_{n\to4} , \lim_{W\to 0}] \delta \la T^{(B)}_{ab}\ra\ .
\eea

Moreover, the identity implies the following $\sigma$ invariant form:
\bea
\label{more}
\alpha~ {\delta\over (n-4) \delta \sigma(x)} \Big[\int d^n x'\sqrt{-g}{
{\cal I}^{(4)}}- \int d^{4} x'\sqrt{-g}{
{I}^{(4)}}-{1\over 8} \int d^4x' (n-4)\sqrt{-g} I^{(4)} \ln \bar g(x')\Big]=0 \ ,\nn\\ 
\eea
that can be freely added into \eqref{new} with an arbitrary coefficient $\alpha$. In total it gives non-zero contribution to the stress tensor after the metric variation as given in \eqref{k}. As before, we should further introduce an $\alpha {1\over 18} R^2$ term that makes the result to the form $(DDW+1/2 RW)$ when combined with the first two terms in \eqref{more}. Note that $\alpha$ will lead to a different coefficient of $D^2 R$ in the trace anomaly. Hence it would change the scheme. Fixing the coefficient of $D^2 R$ under a given scheme is needed to completely fix $\alpha$.

\section{Discussion}

Let us relate this work with \cite{Page1}, where a general (trial) solution to the differential equation \eqref{final} was given by
\bea
\label{page}
&&\la \bar T^\m_\n \ra = \Omega^{-4}\la T^\m_\n\ra -{a_4\over (4\pi)^2}\Big[(4\bar R^{\lambda}_{\rho}\bar W^{\rho\m}_{~~~\lambda\n}-2\bar H^{\m}_{\n})-\Omega^{-4}(4 R^{\lambda}_{\rho}W^{\rho\m}_{~~~\lambda\n}-2 H^{\m}_{\n})\Big]\nn\\
&&-{\gamma\over 6 (4\pi)^2}\Big[I^{\m}_{\n}-\Omega^{-4}I^{\m}_{\n}\Big]-8{c_4\over (4\pi)^2}\Big[\bar D^{\rho}\bar D_{\lambda}(\bar W^{\rho\m}_{~~~\lambda\n}\ln\Omega)+{1\over2} \bar R^{\lambda}_{\rho}\bar W^{\rho\m}_{~~~\lambda\n}\ln\Omega\Big]\ ,
\eea 
where we have expressed it under the same convention defined by \eqref{tracegeneral}. And 
\bea
H_{\m\n} &\equiv&{-{1\over 2}} \Big[g_{\m\n} \Big({R^2\over 2}-R^2_{\lambda \rho} \Big)+2R_{\m}^{\lambda}R_{\n\lambda}-{4\over3}RR_{\m\n}\Big] \ , \\
I_{\m\n} &\equiv&2D_{\m} D_{\n} R-2g_{\m\n}D^2 R-2RR_{\m\n}+{1\over 2}g_{\m\n}R^2\ .
\eea
The corresponding results from the type A and type D anomaly parts agree with the results obtained from the dimensional regularization. The only mismatch part comes from the type B anomaly.  
The following is our explanation, which is again coming from the ambiguity. We note that the result \eqref{page} could be derived by varying the effective action given in $eq(2.2-2.4)$ in \cite{Page2} with respect to the metric. They are in fact the so-called dilaton actions. That is to say, we can re-produce \eqref{page}  by simply adopting these dilaton actions in our formulation. 
However, there might be some potential issues. The first issue is that these dilaton actions were written down with the explicitly given $\sigma$.  One uses these dilaton actions because their $\sigma$ variations give the correct trace anomalies. However, in the context of the dimensional regularization, we see it is certainly not the only way to re-write the anomalies into $\sigma$-exact forms. Allowing the explicit $\sigma$ to appear will generate more ambiguities. Moreover, there is another issue that was already mentioned in \cite{Page2} (in the paragraph between $eq(2.20-2.24)$): They need to impose assumptions on the spacetime in order to deal with the metric variation on the explicit $\sigma$. However, the stress tensors are obtained from the metric variation. If we use the dilaton action, it might lose the spirit of the dimensional regularization where the results are fully expressed as curvature tensor forms instead of working out the $\sigma$'s metric variation.

Finally, we would like to comment briefly on the relation between our present work with the corresponding holographic (AdS/CFT) approach (\cite{Henningson:1998gx},\cite{Balasubramanian:1999re}, \cite{deHaro:2000xn}). The ambiguities were in fact mentioned in \cite{Balasubramanian:1999re} and \cite{deHaro:2000xn} where one can add a local counterterm proportional to the trace anomaly and the coefficient of $D^2 R$ term is arbitrary since it could be generated by adding an $R^2$ term in the action. A gravitational result can be used to match a field theory result only when a scheme is given. In the case of using Einstein gravity, these gravity results are applied to $a_4=c_4$. In particular, in \cite{deHaro:2000xn}, they call the corresponding quantity (defined by the metric variation on 4D/6D trace anomalies) as $h_{(4)}$ for 4D and $h_{(6)}$ for 6D. They ignored these terms from time to time in their paper (refer to eq (3.15) and eq(3.16)) because of the ambiguity. Notice that the stress tensors obtained in \cite{deHaro:2000xn}  without using conformal flatness condition are only formal in the sense that $g_{4}$ in their eq (3.15) and $g_{6}$ in their eq(3.16) are in fact singular (we refer readers to the appendix A in \cite{deHaro:2000xn} for the detailed expressions). Conformal flatness would provide $g_{(4)}={1\over 4} g_{(2)}$ and $g_{(6)}=0$. Hence, one would have finite results. Presumably, a careful further regularization on the gravity side in general backgrounds would allow us to better compare the gravity results with field theory results discussed in this paper. 
\\

{\bf Acknowledgments}: The author is grateful to Christopher Herzog for useful discussion and comments on the manuscript. This work was supported in part by the National Science Foundation under Grant No. PHY-0844827.

\end{document}